\documentclass[Physsubmission, Phys]{SciPost}

\hypersetup{
    colorlinks,
    linkcolor={red!50!black},
    citecolor={blue!50!black},
    urlcolor={blue!80!black}
}

\usepackage[bitstream-charter]{mathdesign}
\DeclareSymbolFont{usualmathcal}{OMS}{cmsy}{m}{n}
\DeclareSymbolFontAlphabet{\mathcal}{usualmathcal}

\begin{document}

\begin{center}{\Large \textbf{
Segal's contractions, AdS and conformal groups\\
}}\end{center}

\begin{center}

{\large Daniel Sternheimer\textsuperscript{1} \textsuperscript{2}  \textsuperscript{3} }
\end{center}

\begin{center}
{\bf 1} Department of Mathematics, Rikkyo University, Tokyo, Japan
\\
{\bf 2} Institut de Math\'ematiques de Bourgogne, Dijon, France,
\\
{\bf 3} Honorary Professor, St.Petersburg State University, and Member of the 
Board of Governors, Ben Gurion University of the Negev, Israel 
\\
\texttt{Daniel.Sternheimer@u-bourgogne.fr}
\end{center}

\begin{center}
\today
\end{center}


\definecolor{palegray}{gray}{0.95}
\begin{center}
\colorbox{palegray}{
  \begin{tabular}{rr}
  \begin{minipage}{0.1\textwidth}
    \includegraphics[width=20mm]{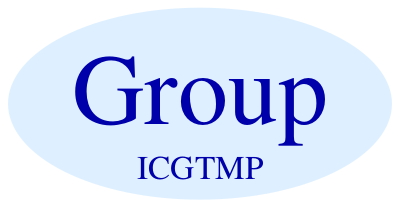}
  \end{minipage}
  &
  \begin{minipage}{0.85\textwidth}
    \begin{center}
    {\it 34th International Colloquium on Group Theoretical Methods in Physics}\\
    {\it Strasbourg, 18-22 July 2022} \\
    \doi{10.21468/SciPostPhysProc.?}\\
    \end{center}
  \end{minipage}
\end{tabular}
}
\end{center}

 \begin{center}
{\bf Dedicated to Irving Ezra Segal (1918-1998) \\
in commemoration of the centenary of his birth.}
\end{center}

\section*{Abstract}
{\bf
Symmetries and their applications always played an important role 
in I.E. Segal's work. I shall exemplify this, starting with his correct proof 
(at the Lie group level) of what physicists call the ``O'Raifeartaigh theorem",
continuing with his incidental introduction in 1951 of the (1953) 
In\"on\"u--Wigner contractions, of which the passage from AdS (SO(2,3)) to 
Poincar\'e is an important example, and with his interest in conformal groups 
in the latter part of last century. Since the 60s Flato and I had many fruitful 
interactions with him around these topics. In a last section I succintly relate 
these interests in symmetries with several of ours, especially elementary 
particles symmetries and deformation quantization, and with an ongoing program 
combining both.
}

\vspace{10pt}
\noindent\rule{\textwidth}{1pt}
\tableofcontents\thispagestyle{fancy}
\noindent\rule{\textwidth}{1pt}
\vspace{10pt}

\section{Prologue}
In July 2018 a special session dedicated to Irving Ezra Segal (13 September 
1918 -- 30 August, 1998) was organized during the first day of the 32$^\textrm{nd}$ 
International Colloquium on Group Theoretical Methods in Physics (Group32) 
that was held at Czech Technical University in Prague, Czech Republic, from 
Monday 9$^\textrm{th}$ July until Friday 13$^\textrm{th}$ July 2018. It was 
meant to be a homage to this immense scientist on the occasion of the centenary 
of his birth and became a commemoration. In the Notices of the American 
Mathematical Society \cite{SegAMS99} were published contributions concerning 
the life and work of I. E. Segal by a number of leading scientists, part of 
whom are/were not with us anymore [Baez, John C.; Beschler, Edwin F.; 
Gross, Leonard; Kostant, Bertram; Nelson, Edward; Vergne, Mich\`ele; 
Wightman, Arthur S.]. Among these I will only quote what Edward Nelson 
(1932 - 2014) said of him (p. 661): \textit{It is rare for a 
mathematician to produce a life work that at the time can be fully and 
confidently evaluated by no one, but the full impact of the work of Irving Ezra 
Segal will become known only to future generations.}  

The text of my invited talk in that special session was sent by me in December 
2018 to the organizers, and sent by them to IOP in early 2019, together with 
all other contributions. In April 2019 the editors informed me that the 
Proceedings of Group32 were published, with a link that recently changed 
to https://iopscience.iop.org/issue/1742-6596/1194/1
(IOP publishes a very large number of conference proceedings, mostly in physics.) 
In December 2021, looking for a more precise reference, I was surprised not to 
find there my contribution. Apparently someone at IOP ``forgot" to include my
contribution, \textit{without informing me nor the organizers of the fact}.  
The organizers of Group34 (in Strasbourg) very kindly agreed to include my text,
which as the reader can see deals indeed with ``Group Theoretical Methods in 
Physics", in the Proceedings of Group34. The sections of the following text are
essentially my original (December 2018) contribution to Group32.

\section{Some history, anecdotes and background material}

\subsection{First interactions with Segal}

The first interactions we (Moshe Flato and I) had with I.E. Segal were probably
on the occasion of the controversy that arose in 1965 around what physicists
still call ``the O'Raifeartaigh theorem". 
Indeed in 1965 Moshe and I submitted to the Physical Review Letters (PRL) a 
contribution \cite{FS65} criticizing that of Lochlainn O'Raifeartaigh, 
published there the same year \cite{OR65}. In the latter paper was ``proved" 
that the so-called ``internal" (unitary) and external (Poincar\'e) symmetries 
of elementary particles can be combined only by direct product. In our rebuttal 
Moshe insisted that we write that the proof (of O'Raifeartaigh, who by the way 
became a good friend after we met) was ``lacking mathematical rigor," a 
qualification which incidentally (especially at that time) many physicists 
might consider as a compliment. Our formulation was deliberately provocative, 
because Moshe felt that we were criticizing a ``result" which, for a variety of 
reasons, many in the ``main stream" wanted to be true. 

{\small {\it{Remark}}. The ``theorem" of O'Raifeartaigh was formulated at the 
Lie algebra level, where the proof is not correct because it implicitly assumes 
that there is a common domain of analytic vectors for all the generators of an 
algebra containing both symmetries. In fact, as it was formulated, the result 
is even wrong, as we exemplified later with counterexamples. The result was 
proved shortly afterward by Res Jost and, independently, by Irving Segal 
\cite{Se67} but only in the more limited context of unitary representations of 
Lie groups. In those days ``elementary particle spectroscopy" was performed 
mimicking what had been done in atomic and molecular spectroscopy, where one 
uses a unitary group of symmetries of the (known) forces. As a student of Racah, 
Moshe mastered these techniques. The latter approach was extended somehow to 
nuclear physics, then to particle physics. That is how, to distinguish between 
neutrons and protons, Heisenberg introduced in 1932 ``isospin," with SU(2) 
symmetry. When ``strange" particles were discovered in the 50s, it became 
natural to try and use as ``internal" symmetry a rank-2 compact Lie group. 
In early 1961 Fronsdal and Ben Lee, with Behrends and Dreitlein, all present 
then at UPenn, studied all of these. At the same time Salam asked his PhD 
student Ne'eman to study only SU(3), in what was then coined ``the Eight-Fold 
way" by Gell'Mann because its eight-dimensional adjoint representation could be 
associated with mesons of spin 0 and 1, and baryons of spin $\frac{1}{2}$. Since 
spin is a property associated with the ``external" Poincar\'e group, it was 
simpler to assume that the two are related by direct product. Hence the interest 
in the ``O'Raifeartaigh theorem." For this and much more see e.g. Section~2 in 
\cite{GMP32} and references therein.
The Editors of PRL objected to our formulation. In line with the famous Einstein 
quote (``The important thing is not to stop questioning, curiosity has its 
reason for existing.") Moshe insisted on keeping it ``as is." The matter went up 
to the President of the American Physical Society, who at that time was Felix 
Bloch, who consulted his close friend Isidor Rabi. [In short, Rabi discovered 
NMR, which is at the base of MRI, due to Bloch.] Rabi naturally asked who is 
insisting that much. When he learned that it was Moshe, who a few years earlier, 
when Rabi was giving a trimester course in Jerusalem, kept asking hard questions 
which he often could not answer, he said: ``If he insists he must have good 
reasons for it. Do as he wishes." The Editors of PRL followed his advice.}

\subsection{ICM 1966 and around}

The following year (in April 1966), at a conference in Gif-sur-Yvette on 
``the extension of the Poincar\'e group to the internal symmetries of elementary 
particles" which Moshe (then 29) naturally co-organized, Christian Fronsdal told 
Moshe: ``You wrote that impolite paper." This was the beginning of a long 
friendship, which lasts to this day and is at the origin of important scientific 
works, many of which deal with applications of group theory in physics, and 
related issues on quantization. 

Shortly afterward (16--26 August 1966) an important mathematical event happened:
the International Congress of Mathematicians (where the acronym ICM came into 
wide usage). Until then the scientific exchanges between the USSR and ``the West" 
had been very limited. A record number of mathematicians attended (4,282 
according to official statistics), of which 1,479 came from the USSR, 672 from 
other ``Socialist countries" in Europe, while over 1,200 came from ``Western 
countries", including 280 from France: Moshe (then still only citizen of Israel 
but working in France since October 1963) and I were among the latter. Irving 
Segal came from the US. I remember that we and many from the French delegation
traveled to Moscow on a Tupolev plane, organized as in a train with compartments
seating eight. In Moscow, we were accommodated, together with many ``ordinary" 
participants (some VIPs, among them Segal if I remember correctly, were 
accommodated in ``smaller" hotels closer to the Kremlin), in the huge hotel 
Ukraina (opened in 1957, the largest hotel in Europe), one of the seven 
Stalinist skyscrapers in Moscow with a height of 206 meters (including the 
spire, 73 meters long) and total floor area ca. 88,000m$^2$, a small city 
in itself. On every floor there were ``etazhniks"  supposed
to help but in fact checking on the guests. (We encountered the same 
system 10 years later in Taipei . . . ).

The opening ceremony of ICM 1966, as it is now known, was held in the Kremlin.
During the long party which followed we met, and instantly became friends 
with, many leading Soviet scientists like Nikolay Nikolayevich Bogolyubov 
(who invited us to Dubna after the Congress), Ludwig Dmitrievich Faddeev and 
Israel Moiseevich Gelfand. There were also a number of cultural events, both 
official and optional. I remember that, near a centrally located hotel, 
we (Moshe and I) and Segal were looking for a taxi in order to get to one such 
event. Segal entered in a random run, trying to get one, to no avail. Moshe, 
who spoke fluent Russian (though he could not read nor write it), calmly managed 
with the doorman of the hotel to get one for the three of us.  

At the end of 1966 we made our first visit to the US, starting with Princeton 
at the invitation of E.P. Wigner with whom Moshe (being a student of Racah) had 
established connection. We also went to Brookhaven National Laboratory (where 
my cousin Rudolph Sternheimer spent most of his career, and where the editor 
of PRL, Sam Goudsmith, was located). Very generously Segal invited us to MIT, 
then and during our following visits (in 1969 and later) and accommodated us 
in the Sheraton Commander near Harvard square. 

We had many subjects of common interest, mostly around group theory in relation
with physics (for some, see below). On the anecdotical side, at some point 
during our second visit to US (in 1969, with Jacques Simon) the discussion came 
around Fock space \cite{Fo32}. Segal insisted that it should be called 
``Fock -- Cook space" because his student Joseph M.Cook made it more precise 
in his 1951 Thesis in Chicago (ProQuest Dissertations Publishing, 1951. T-01196).  
Not surprisingly that unusual terminology did not catch.  
[F.J. Dyson wrote in Mathematical Reviews, about the announcement in PNAS: 
``The author has set up a mathematically precise and rigorous formulation of 
the theory of a linear quantized field, avoiding the use of singular functions.
The formalism is equivalent to the usual one, only it is more carefully 
constructed, so that every operator is a well-defined Hilbert space operator 
and every equation has an unambiguous meaning. Fields obeying either Fermi or 
Bose statistics are included."]

\subsection{Later interactions.}

An anecdotical event, among our later interactions, also related to the Soviet 
Union (of the latter days) is the following. Our first visit to Leningrad (also 
technically the last one, because our subsequent visits were to 
St. Petersburg ...) occured in the Spring of 1989. We (Moshe and I) were 
accommodated in a recent Finnish-built hotel, not very fancy and at the entrance 
of the city when coming from the airport, but functional. Alain Connes, who 
visited at the same time, had a nice room in a top floor of Hotel Evropeiskaya 
(now back to its original splendor and named Grand Hotel Europe); however at the 
time the hotel was rather run down and e.g. water reached his room only
a few hours per day! 

If I remember correctly it was then that we met in the USSR another visitor, 
Irving Segal, who introduced us to his second wife Martha Fox, whom he had 
married in 1985. [His first wife, Osa Skotting, had left him at some point 
(some said for another woman) and not long afterward remarried in 1986 with an 
old flame of her, Saunders Mac Lane, whom we met on many occasions in the 
University of Chicago, always dressed in tartan trousers (the MacLean tartan, 
of course).] 

Interestingly we met both Irving and Martha shortly thereafter at a workshop 
in Varna, where we were all accommodated in a nice ``rest house" for the 
``nomenklatura" of the Bulgarian Communist party. There we met also for the 
first time Vladimir Drinfeld and a number of other ``Eastern bloc" 
mathematicians. We were warned by our Bulgarian friend Ivan Todorov (who, as
physicist and Academician, had access to information that was not widely 
publicized) to be careful when drinking wine, because after the Chernobyl 
disaster in April 1986, many agricultural products (including mushrooms and 
especially wines) produced in the following months and years were contaminated. 
[At that time French authorities claimed that the Chernobyl radioactive cloud 
did not cross the Rhine, which of course nobody believed.] In any case there 
were still enough older wines in Bulgaria for us to enjoy in the evenings, 
and we did, including with Martha who liked the company of these (then younger) 
scientists. One evening Irving spent some time in scientific discussions with 
a distinguished colleague after which, seeing us in the lobby, he said: 
``Martha you are tired, please come". Martha denied being tired but then 
Irving insisted:``Martha you are tired, and besides you have some duties 
to perform." At this point Joe Wolf quipped: ``Not here I hope!" 
She had to leave. I told the story later to some friends in the US, and one of 
them remarked: ``It must have worked because recently at MIT Segal has been 
distributing cigars on the occasion of the birth of their daughter Miriam."

\section{Contractions, conformal group \& covariant equations}

\subsection{A tachyonic survey of contractions and related notions}
In 1951, in a side remark at the end of an article \cite{Se51}, Segal introduced
the notion of contractions of Lie algebras, that was ``introduced in physics"
in a more explicit form two years later \cite{IW53} by Eugene Wigner and Erdal 
In\"on\"u. [The latter was the son of Ismet In\"on\"u who in 1938 succeeded 
Ataturk as president of Turkey. Eventually Erdal (1926 -- 2007) had a political 
career, becoming himself interim Prime Minister in 1993.] The notion has been 
studied and generalized by a number of people. For an informative more recent 
paper, see e.g. \cite{WW00}. (I had something to do with its publication.)

The 1951 paper by Segal was analyzed in Mathematical Reviews by Roger Godememt.
It included some nasty remarks (something rare in the Reviews but not infrequent 
with Godement), in particular, after giving a number of simpler proofs of a 
few results, Godement wrote:
``Tout cela est tr\`es facile. L'article se termine par quelques exemples 
inspir\'es de probl\`emes physiques, \`a propos desquels l'auteur \'emet des 
opinions et suggestions dont la discussion demanderait des connaissances 
cosmologiques et m\'etaphysiques que le rapporteur n'a malheureusement pas eu 
le temps d'acqu\'erir." {\small{[All this is very easy. The paper ends with 
some examples inspired by physical problems; in connection with these the 
author expresses opinions and suggestions, the discussion of which would 
require cosmological and metaphysical knowlegde which the reviewer 
unfortunately did not have the time to obtain.]}} 
These examples include the notion of contractions of Lie algebras, and more!

In a nutshell, a typical example of contraction consists in multiplying part of
the generators of some linear basis of a Lie algebra by a parameter $\epsilon$
which then is made 0. In particular, when multiplying the Lorentz boosts of
the Lie algebra $\mathfrak{so}(3,1)$ by $\epsilon \to 0$, one obtains the
Lie algebra of the Euclidean group $E(3)$ ($\mathfrak{so}(3) \cdot \mathbb{R}^3$).
   
The notion of contractions of Lie algebras is a kind of inverse of the more
precise notion that became known, ten years later, after the seminal paper by 
Murray Gerstenhaber \cite{Ge64}, as deformations of (Lie) algebras. 
Immediately thereafter it became clear to many, especially in France where
Moshe Flato had arrived in 1963, that the symmetry of special relativity
(the Poincar\'e Lie algebra $\mathfrak{so}(3,1) \cdot \mathbb{R}^4$) is a 
deformation of that of Newtonian mechanics (the Galilean Lie algebra, 
semi-direct product of $E(3)$ in which the $\mathbb{R}^3$ are velocity 
translations, and of space-time translations). Or, conversely, that Newtonian 
mechanics is a contraction (in the sense of Segal), of special relativity. 

These notions were extensively discussed at the above-mentioned April 1966 
conference in Gif. In other words, special relativity can be viewed, from the 
symmetries viewpoint, as a deformation. A natural question, which already then 
arose in the mind of Moshe, was then to ask whether quantum mechanics, the 
other major physical discovery of the first half of last century, can also be 
viewed as a deformation. It was more or less felt, because of the notion of 
``classical limit" and though in this case we deal with an infinite dimensional 
Lie algebra, that classical mechanics is a kind of contraction (when 
$\hbar \to 0$) of the more elaborate notion of quantum mechanics. But the 
inverse operation is far from obvious, if only because in quantum mechanics 
the bracket is the commutator of operators on some Hilbert space while in 
classical mechanics we deal with the Poisson bracket of classical observables, 
functions on some phase space.     

At the same time, in 1963/64, I participated in the Cartan--Schwartz seminar
at IHP (Institut Henri Poincar\'e) on the (proof of the) seminal theorem
of Atiyah and Singer on the index of elliptic operators, which had just been 
announced without a proof. My share (2 talks), on the Schwartz side, in the 
Spring of 1964, was the multiplicative property of the analytic index, crucial
for achieving dimensional reduction that was an important ingredient of the
proof. Moshe followed that important seminar. But it was only more than a 
dozen years later, after we developed what became known as deformation 
quantization, that we realized that the composition of symbols of differential 
operators was a deformation of the commutative product of functions (what we 
called a star-product), or conversely that the commutative product is a 
contraction of our star-product.           

\subsection{Conformal groups and conformally covariant equations}

The conformal group (here, the Lie group $SO(4,2)$ or a covering of it) was 
introduced ten years before Segal was born as a symmetry of Maxwell equations
by Harry Bateman (in 1908 and 1910) \cite{Ba08,Ba10} and Ebenezer Cunningham 
(in 1910) \cite{Cu10}. In 1936 (a year after an article, also in Ann. Math., 
in which he studied the extension of the electron wave equation to de Sitter 
spaces) Dirac \cite{Di36} made this fact more precise. Yet not many realized 
the fact, possibly because in addition to the Poincar\'e group (a group of linear 
transformations of space-time) there were 4 generators of ``inversions", 
nonlinear transformations. For a long time many (including textbooks authors 
and some colleagues physicists of Moshe in Dijon) were convinced that the 
Poincar\'e group is the most general group of invariance of special relativity. 

We were introduced to this group in 1965 by Roger Penrose during a visit to 
David Bohm at Birbeck College of the university of London. We were impressed 
by Roger Penrose. When we told that to Andr\'e Lichnerowicz he remarked: 
``Half of what he says is true." That half proved to be seminal and worth a 
Nobel prize. Inasmuch as the conformal group is concerned we made immediate use 
of it in a number of papers in a variety of contexts \cite{BFSV, FS66}. 

In December 1969 Moshe was visiting KTH (the Royal Institute of Technology)
in Stockholm, at the time when Gell'mann gave there the traditional scientific 
lecture on the occasion of his Nobel prize. For an unknown reason, he chose to 
center his talk on the conformal group, which by then we knew very well, 
in particular because we had studied in detail the conformal covariance of 
field equations \cite{FSS70}. Moshe (then 32) did not hesitate to interrupt 
him a few times, asking from the back of the auditorium (im)pertinent questions 
to which Gell'Mann's only answer was: ``Good question." Eventually Moshe said 
that he did not ask for marks for his questions, but would like answers. 
At that point Gell'Mann, who is known to be very fast, remarked: ``I didn't 
know there would be specialists in the audience." Then Moshe, who was even 
faster, replied: ``Until now you insulted only me, now you are insulting the 
Nobel Committee, who is sitting here [in the first row]." That is not a good 
way to make friends. After the lecture half of the Committee members, instead 
of joining Gell'Mann for a lunch at the US Embassy, joined Moshe for a (better) 
lunch at the French Embassy, in honor of Samuel Beckett who that year was the 
Nobel laureate in Literature (but had sent his publisher to collect the Prize, 
prefering to remain in sunny Tunisia with his young companions). 

For some more information on Moshe Flato (who coincidentally was born September 
17, 4 days and 19 years after Segal, and, like Segal, died in 1998, almost 3 
months after Segal) see e.g. \cite{CMF99,MFre99}.    

In addition to the above mentioned papers, we published a few other papers 
around the conformal group. Moshe had discussed the issue with Segal, who at 
first didn't seem interested in the idea. But the question apparently remained 
in his mind and not long afterward he dealt with the conformal group from a 
different point of view. In his first of many publications on the subject 
\cite{Se71} (reviewed by Victor Guillemin) and \cite{Se76} appears the 
universal covering space of the conformal compactification of Minkowski space, 
in connection with a simple explanation for the ``red shift" observed by 
astronomers in studying quasars. At the time, though he had many more important 
contributions (albeit mostly of mathematical nature) Segal was very proud of 
his explanation, in spite of the fact that many astronomers were critical, 
because while his explanation worked well for some galaxies, it did not work 
so well for others, which Segal did not consider. That is one more example of 
what Sir Michael Atiyah said at the International Congress of Mathematical 
Physics in London in 2000 (his contribution there was published in \cite{Ati00}): 
``Mathematicians and physicists are two communities separated by a common 
language."  This refers to a famous saying which is often attributed to George 
Bernard Shaw but seems to date back to Oscar Wilde (in ``The Canterville Ghost", 
1887): ``We have really everything in common with America nowadays, except, of 
course, language."

\subsection{Relativistically Covariant Equations}

In 1960 Segal published an important paper \cite{Se60} in the first volume of 
the Journal of Mathematical Physics, extensively (often with personal remarks, 
e.g. in relation with QED) analyzed in Mathematical Reviews by Arthur Wightman, 
in a review much longer than the abstract of the paper. Among many interesting 
ideas, that were further developed in subsequent papers, appears there a 
Poisson bracket on the (infinite dimensional) space of initial conditions for 
the Klein Gordon equation.

Our interest in that structure was triggered by the approach we made, from the 
70s to the 90s and in parallel with deformation quantization (see below), of
many nonlinear evolution equations of physics, as covariant under a kind of
deformation of the symmetry of linear (free) part. [That approach has not yet
attracted enough attention from specialists of these PDEs and ODEs, possibly
because the tools used, involving e.g. group representations and their 
cohomologies, which serve as a basis for a careful analytical study of such 
equations, are foreign to PDEs specialists.] It culminated in the ``tour de 
force" extensive \cite{FST97} study of CED (classical electrodynamics), namely 
``Asymptotic completeness, global existence and the infrared problem for 
the Maxwell-Dirac equations" (see also references therein and a few later 
developments), where it is explained in detail. 

Very appropriately it is dedicated to the memory of Julian Schwinger, 
``the chief creator of QED" (certainly in its analytical form). Indeed a 
rigorous passage from CED to QED, from the point of view of deformation
quantization, will require a Hamiltonian structure on the space of initial
conditions for CED, of the kind introduced by Segal, the quantized fields 
being considered as functionals on that space. That is one more example of
how our works were, and should be in the future, intimately intertwined with 
those of Segal.              

\section{AdS, AdS/CFT, deformation quantization \& perspectives}

In this section I succintly present my ongoing research on the convergence 
of topics which we were concerned with in the 60s, around 
symmetries of elementary particles, with later works (from the 70s) on the 
essence of quantization and around conformal groups. The former are closely 
related to our above mentioned first interaction with Segal and the latter 
to applications in physics of mathematical tools he developed, in relation 
with both quantization and conformal groups. An early presentation can be found
e.g. in \cite{GMP32}.  

\subsection{Classical limit, deformation quantization and avatars}

Since that part is developed in numerous reviews, by many, I shall here give
only an ultra-short presentation, starting with the connection with Segal's 
contractions. Indeed, the fact that classical mechanics is, in a sense, 
a ``contraction" of quantum mechanics, was essentially known to many, one of 
the first being Dirac \cite{Di33}, and has been expressed precisely e.g. 
by Hepp in \cite{He74}. Quite naturally the idea that quantization should be 
some kind of a deformation was ``in the back of the mind" of many, but how to
express that precisely was far from obvious. After \cite{BFFLS1,BFFLS2} appeared 
one of them demanded from Andr\'e Lichnerowicz to be quoted for the idea, but 
Andr\'e did not know how we could include ``the back of the mind" of that 
person in our list of references! [Incidentally our 1977 UCLA preprint of 
\cite{BFFLS1,BFFLS2} was sent to Annals of Physics by Schwinger, who had 
published in 1960 \cite{JS60} a short paper which turned out to be related to it.]

That it should be possible to formulate such an idea in a mathematically precise 
way was implicitly felt by Dirac in \cite{Di51}, where he went on by developing
his approach to quantization of constrained systems (in geometrical language,
coupled second class constraints reduces $\mathbb{R}^{2n}$ phase space to a
symplectic submanifold, and first class constraints reduce it further to what
we called a Poisson manifold):

{\small{... One should examine closely even the elementary and the satisfactory 
features of our Quantum Mechanics and criticize them and try to modify them, 
because there may still be faults in them. The only way in which one can hope
to proceed on those lines is by looking at the basic features of our present 
Quantum Theory from all possible points of view. Two points of view may be 
mathematically equivalent and you may think for that reason if you understand 
one of them you need not bother about the other and can neglect it. But it 
may be that one point of view may suggest a future development which another
point does not suggest, and although in their present state the two points of 
view are equivalent they may lead to different possibilities for the future. 
Therefore, I think that we cannot afford to neglect any possible point of
view for looking at Quantum Mechanics and in particular its relation to 
Classical Mechanics. Any point of view which gives us any interesting feature 
and any novel idea should be closely examined to see whether they suggest any
modification or any way of developing the theory along new lines. ...}}      
 
That is the path we followed in our foundational papers \cite{BFFLS1,BFFLS2} 
that are extensively quoted, directly and even more implicitly. The notion became 
a classic, and constitutes an item in the Mathematics Subject Classification. 
For a detailed review see e.g. \cite{DS02}. The even more developed notions of 
quantum groups and of noncommutative geometry, which had different origins, 
appeared essentially shortly afterward and may be considered as avatars. 
For this and more, see e.g. \cite{DS11,GMP32}.

\subsection{AdS, AdS/CFT and particle physics}
As is well known, the Anti de Sitter group AdS$_4$, $SO(2,3)$ (or a covering 
of it), can be viewed either as the conformal group of a 2+1 dimensional flat
space-time, or a deformation (with negative curvature) of the Poincar\'e 
group of usual Minkowski (3+1) dimensional space-time, the latter being a
``Segal" contraction of AdS$_4$. That has many physical consequences, including
for particle physics, which have been studied by many authors, especially since
the 70s. See e.g. \cite{AF$^2$S81} (which has been quoted by Witten as an early 
instance of the AdS/CFT correspondence) and references therein, where e.g. is
shown how AdS representations contract to the Poincar\'e group, and many later 
papers by us and others. 

The two massless representations of the Poincar\'e group in 2+1 dimensions have
a unique extension to representations of its conformal group AdS$_4$ (that 
feature exists in any higher dimension). The latter were discovered in 1963 by 
Dirac and called by him ``singletons". We called them ``Rac" for the scalar 
perticle (because it has only one component, ``Rac" means only in Hebrew) and
``Di" for the helicity $\frac{1}{2}$ one (which has 2 components), on the 
pattern of Dirac's ``bra" and ``ket".

Among the many applications to particle physics, including with conformal 
symmetry, one should mention that the photon can be considered as 
\emph{dynamically} composed of two Racs, in a way compatible with QED 
\cite{FF88}, and that the leptons can also be considered as composites of 
singletons \cite{CF00}, in a way generalizing the electroweak unification
theory. Thus, in the same way as (special) relativity and quantum mechanics
can be considered as deformations, deforming Minkowski space to Anti de Sitter
(with a tiny negative curvature) can explain photons and leptons as composites.
A natural question is how to extend that to the heavier hadrons. 

The approach I am advocating (\cite{GMP32} and work in progress, in particular 
a Springer Brief in Mathematical Physics with Milen Yakimov), based on the 
strong belief that one passes from one level of physical theories to another 
by a deformation in some category, is to deform the symmetry one step further, 
to some quantized Anti de Sitter (qAdS), possibly with multiple parameters 
(commuting so far, since one does not know yet how to do treat deformations 
with noncommutative parameters, e.g. quaternions), and even at roots of unity 
since the Hopf algebra of quantum groups at roots of unity is finite dimensional. 
Maybe one could then find the ``internal symmetries" as symmetries of the 
deformation ``parameters", putting on solid ground that ``colossus with clay 
feet" called the Standard Model. Vast programme, as could have said de Gaulle. 
Interestingly the development of the required mathematics (of independent 
interest) would be related to a number of Segal's works.  

\section{Conclusion}
The above scientific and anecdotical samples show how, in spite of being 
a generation apart, our lives have been ``intertwined" for over 30 years with
that of I.E. Segal. The use of symmetries in physics have been a kind of 
watermark throughout our works, beyond their apparent diversity. A special 
mention is due to the conformal group (of Minkowski space-time) which has played
an important role throughout the works of I.E. Segal, in particular in his late
cosmological applications. In this century, very modestly, I have been trying 
to develop (unconventional in a different way) consequences in that direction.  

\section*{Acknowledgements}
First I want to thank the organizers of Group32, in particular Patrick Moylan
who managed the special session dedicated to I.E. Segal, for inviting me to it.
Very special thanks are due to the organizers of 34$^\textrm{th}$ International 
Colloquium on Group Theoretical Methods in Physics (Group34), in particular to 
the Chair of its Organizing Committee Michel Rausch de Traubenberg, for kindly
giving ``hospitality" to my contribution in honor of I.E. Segal, among those of
Group34, which are being published with SciPost. [They also invited me to 
attend the meeting in Strasbourg, which I could not accept, inter alia, because 
of the last wave of Covid-19.] And to Rutwig Campoamor-Stursberg, a 
distinguished member of the Group34 Organizing Committee, for the ``TeXploit" 
of adapting my plain LaTeX contribution to the (unfriendly 
for me, as was the rigid IOP style) requirements of the SciPost style, 
including concerning the bibliography.   





 \nolinenumbers


\begin{thebibliography}{99}


\bibitem[AF$^2$S81]{AF$^2$S81}
E. Angelopoulos, M. Flato, C. Fronsdal, and D. Sternheimer, 
\emph{Massless particles, conformal group, and de Sitter universe}, Phys. Rev. D (3) 
\textbf{23}, no. 6, 1278 (1981); \doi{10.1103/PhysRevD.23.1278}.

\bibitem[Ati00]{Ati00}
M. Atiyah, \emph{On the unreasonable effectiveness of physics in 
mathematics.} Highlights of mathematical physics (London, 2000), 25--38, 
Amer. Math. Soc., Providence, RI, 2002. 

\bibitem[Ba08]{Ba08}
H. Bateman, \emph{The conformal transformations of a space of four dimensions
and their applications to geometrical optics}, Proc. London Math. Society 
\textbf{7}, 70 (1908); \doi{10.1112/plms/s2-7.1.70}.

\bibitem[Ba10]{Ba10}
H. Bateman, \emph{The Transformation of the Electrodynamical Equations}, 
Proc. London Math. Society \textbf{8}, 223 (1910); \doi{10.1112/plms/s2-8.1.223}.

\bibitem[BFFLS1]{BFFLS1}
F. Bayen, M. Flato, C. Fronsdal, A. Lichnerowicz and D. Sternheimer, 
\emph{Deformation theory and quantization I. Deformations of symplectic 
structures}. {Ann. Physics} \textbf{111}, 61 (1978); 
\doi{10.1016/0003-4916(78)90224-5}.

\bibitem[BFFLS2]{BFFLS2}
F. Bayen, M. Flato, C. Fronsdal, A. Lichnerowicz and D. Sternheimer, 
\emph{Deformation theory and quantization II. Physical applications}, 
{Ann. Physics} \textbf{111}, 111 (1978); \doi{10.1016/0003-4916(78)90225-7}.

\bibitem[BFSV]{BFSV}
D. Bohm, M. Flato, D. Sternheimer and J.P. Vigier, \emph{Conformal-group 
symmetry of elementary particles}, Nuovo Cimento (10) \textbf{38} 1941 (1965);
\doi{10.1007/BF02750118}.

\bibitem[Co51]{Co51}
J. M. Cook, \emph{The mathematics of second quantization}, 
Trans. Amer. Math. Soc. \textbf{74}, 222  (1953); \doi{10.1073/pnas.37.7.417}

\bibitem[Cu10]{Cu10}
E. Cunningham, \emph{The principle of Relativity in Electrodynamics and 
an Extension Thereof,} Proc. London Math. Soc. \textbf{8}, 77 (1910); 
\doi{10.1112/plms/s2-8.1.77}.

\bibitem[Di33]{Di33}
P. A. M. Dirac, \emph{The Lagrangian in Quantum Mechanics}, Phys. 
Zeitschr. Sowjetunion \textbf{3}, 64 (1933).

 \bibitem[Di36]{Di36}
P. A. M. Dirac, \emph{Wave equations in conformal space}, Ann. Math.  (2), 
\textbf{37}, 429 (1936); \doi{10.2307/1968455}

\bibitem[Di51]{Di51}
P. A. M. Dirac, \emph{The relation of classical to quantum mechanics}, Proc. 
Second Canadian Math. Congress, Vancouver, 1949, pp. 10--31. University of 
Toronto Press (1951). 

\bibitem[DS02]{DS02}
G. Dito and D. Sternheimer, \emph{Deformation quantization: genesis, 
developments and metamorphoses}, in Deformation Quantization (Strasbourg, 2001), 
9--54, IRMA Lect. Math. Theor. Phys., \textbf{1}, de Gruyter, Berlin, 2002; 
\doi{10.1515/9783110866223.9}.

\bibitem[He74]{He74} 
K. Hepp, \emph{The Classical Limit for Quantum Mechanical Correlation 
Functions}, Commun. Math. Phys. \textbf{35}, 265 (1974); \doi{10.1007/BF01646348}

\bibitem[CMF99]{CMF99}
G. Dito and D. Sternheimer (Eds.) \emph{Mosh\'e Flato, the man and the scientist}. 
Conf\'erence Mosh\'e Flato 1999, Vol. I (Dijon), 3--54, Math. Phys. Stud., 
\textbf{21}, Kluwer Acad. Publ., Dordrecht, 2000. ISBN: 0792365402  

\bibitem[MFre99]{MFre99}
\emph{Mosh\'e Flato -- Personal Recollections}, Lett. Math. Phys. 
\textbf{48}, 5 (1999); \doi{ 10.1023/A:1017293905798}

\bibitem[FF88]{FF88}
M. Flato and C. Fronsdal, \emph{Composite electrodynamics}, J. Geom. Phys. 
\textbf{5},  37 (1988); \doi{ 10.1016/0393-0440(88)90013-7}. 

\bibitem[FSS70]{FSS70}
M. Flato, J. Simon and D. Sternheimer, \emph{Conformal covariance 
of field equations}, Ann. Physics \textbf{61} 78 (1970); 
\doi{10.1016/0003-4916(70)90377-5}.

\bibitem[FST97]{FST97}
M. Flato, J. C. H. Simon and E. Taflin, \emph{Asymptotic 
completeness, global existence and the infrared problem for the Maxwell-Dirac 
equations}, Mem. Amer. Math. Soc. \textbf{127}, no. 606, (x+311 pp., 1997). 

\bibitem[FS65]{FS65}
M. Flato and D. Sternheimer, \emph{Remarks on the connection between external
and internal symmetries}, Phys. Rev. Letters \textbf{15}, 934 (1965); 
\doi{10.1103/PhysRevLett.15.934}.

\bibitem[FS66]{FS66}
M. Flato and D. Sternheimer, \emph{Remarques sur les automorphismes 
causals de l'espace-temps}, C. R. Acad. Sci. Paris S\'er. A-B \textbf{263}, 
A953 (1966).

\bibitem[CF00]{CF00}
Ch. Fr\o nsdal, \emph{Singletons and neutrinos}, Lett. Math. Phys. 
\textbf{52}, 51 (2000), \doi{10.1023/A:1007693518414}.

\bibitem[Fo32]{Fo32}
V. Fock, \emph{Konfigurationsraum und zweite Quantelung}, Z. Physik \textbf{75}, 
622  (1932); \doi{10.1007/BF01344458}.

\bibitem[Ge64]{Ge64}
M. Gerstenhaber, \emph{On the deformation of rings and algebras}, 
Ann. of Math. (2) \textbf{79}, 59 (1964); \doi{10.2307/1970484}.

\bibitem[IW53]{IW53}
E. In\"on\"u and E. P. Wigner, \emph{On the contraction of groups and their 
representations}, Proc. Nat. Acad. Sci. U. S. A. \textbf{39}, 510 (1953); 
\doi{10.1073/pnas.39.6.510}.

\bibitem[OR65]{OR65}
L. O'Raifeartaigh, \emph{Mass differences and Lie algebras of finite order},
Phys. Rev. Lett. \textbf{14}, 575 (1965), \doi{10.1103/PhysRevLett.14.575}.

\bibitem[JS60]{JS60}
J. Schwinger, \emph{The special canonical group}, Proc. Nat. Acad. Sci. 
U.S.A. \textbf{46}, 1401 (1960); \doi{10.1073/pnas.46.10.1401}.

\bibitem[Se51]{Se51}
I. E. Segal, \emph{A class of operator algebras which are determined by groups}, 
Duke Math. J. \textbf{18}, 221 (1951); \doi{10.1215/S0012-7094-51-01817-0}. 

\bibitem[Se60]{Se60}
I. E. Segal, \emph{Quantization of nonlinear systems}, J. Mathematical Phys. 
\textbf{1}, 468  (1960); \doi{10.1063/1.1703683}.

\bibitem[Se67]{Se67} 
I. Segal. \emph{An extension of a theorem of L.O'Raifeartaigh.} 
J. Funct. Anal. \textbf{1}, 1 (1967); \doi{10.1016/0022-1236(67)90023-7}.

\bibitem[Se71]{Se71}
I. Segal, \emph{Causally oriented manifolds and groups}, Bull. Amer. Math. 
Soc. \textbf{77} 958  (1971); \doi{10.1090/S0002-9904-1971-12815-X }.

\bibitem[Se76]{Se76}
I. E. Segal, \emph{Theoretical foundations of the chronometric cosmology}, 
Proc. Nat. Acad. Sci. U.S.A. \textbf{73} 669 (1976); \doi{10.1073/pnas.73.3.669}.

\bibitem[SegAMS99]{SegAMS99}
I. E. Segal \emph{Obituary} Notices AMS \textbf{46}(6) 659--668 (2019).
https://www.ams.org/notices/199906/mem-segal.pdf 

\bibitem[DS11]{DS11}
D. Sternheimer, \emph{The deformation philosophy, quantization and 
noncommutative space-time structures}, In: Cattaneo, A., Giaquinto, A., Xu, P. 
(eds) Higher Structures in Geometry and Physics. Progress in Mathematics, 
vol 287. Birkh\"auser, Boston, MA;  \doi{10.1007/978-0-8176-4735-3_3}.

\bibitem[GMP32]{GMP32}
D. Sternheimer, \emph{``The important thing is not to stop questioning", 
including the symmetries on which is based the Standard Model}. 
In: Kielanowski, P., Bieliavsky, P., Odesskii, A., Odzijewicz, A., 
Schlichenmaier, M., Voronov, T. (eds) Geometric Methods in Physics. Trends in 
Mathematics. Birkh\"auser, Cham; \doi{10.1007/978-3-319-06248-8_2}

\bibitem[WW00]{WW00}
E. Weimar-Woods, \emph{Contractions, generalized In\"on\"u--Wigner 
contractions and deformations of finite-dimensional Lie algebras}, 
Rev. Math. Phys. \textbf{12}, 1505 (2000); \doi{10.1142/S0129055X00000605}.

\end{thebibliography}
\end{document}